\documentstyle[11pt,epsfig]{article}
\def\thefootnote{\fnsymbol{footnote}}

\def\be{\begin{equation}}
\def\ee{\end{equation}}
\def\bea{\begin{eqnarray}}
\def\eea{\end{eqnarray}}
\def\ben{\begin{enumerate}}
\def\een{\end{enumerate}}

\def\del{\partial}
\def\np {Nucl. Phys.}

\def\text#1{{\rm #1}}

\def\Lag{{\cal L}}

\def\nlo#1{\mbox{${\cal O}\left(Q^{#1}\right)$}}
\def\mpi{m_\pi}
\def\mpis{{m_\pi^*}}
\def\order#1{{\mbox{${\cal O}\left(Q^{#1}\right)$}}}

\begin{document}
\begin{center}
\begin{flushright}
SNUTP 99-053,
USC(NT)-01-03
\end{flushright}
\vskip 1.0cm
{\Large\bf In-Medium Effective Pion Mass
  \\ from Heavy-Baryon Chiral Perturbation Theory}
\vskip 1cm

{\large Tae-Sun Park$^{a,b,}$\footnote{E-mail:park@nuc003.psc.sc.edu.}
Hong Jung$^{c,}$\footnote{E-mail:jung@sookmyung.ac.kr}
and Dong-Pil Min$^{d,}$\footnote{E-mail:dpmin@phya.snu.ac.kr}}
\vskip 5mm
{\large {\it $^{a}$Theory group, TRIUMF, Vancouver, B. C., Canada V6T 2A3}}
\vskip 1mm
{\large {\it $^{b}$Department of Physics and Astronomy,
University of South Carolina,\\
Columbia SC 29208, USA}}
\vskip 1mm
{\large {\it $^{c}$Department of Physics,
                   Sookmyung Women's University, Seoul 140-742, Korea}}
\vskip 1mm {\large {\it $^{d}$Department of Physics,
                   Seoul National University, Seoul 151-742, Korea}}
\end{center}

\vskip 1cm Using heavy-baryon chiral perturbation theory, we have
calculated all the diagrams up to two-loop order which contribute
to the $S$-wave pion self-energy in symmetric nuclear matter. Some
subtleties related to the definition of pion fields are discussed.
The in-medium pion mass turns out to increase by only $(5
\sim 6)$ per cents at normal nuclear matter density, 
and this result is model independent.

\renewcommand{\thefootnote}{\#\arabic{footnote}}
\setcounter{footnote}{0}

\newpage

The role of pionic interactions in nuclear medium has long been
an important issue in nuclear physics.
Recently,  the  interest in the $S$-wave pionic interaction in nuclear medium
has reemerged due to the experimental
observation of deeply bound pionic states in Pb at
GSI \cite{mpiexp,WBW,FG}.
It is to be noted that
there is a wide variance
among the theoretically deduced values of
the effective pion mass $m_{\pi}^*$;
for symmetric nuclear matter,
$m_{\pi}^*$ is found to increase by 7 \% in Ref.\cite{WBW},
but by 20 \% in Ref.\cite{FG}.
These results are obtained relying on the phenomenology
related to the $S$-wave pion-nucleus optical potential.
It is well-known that threshold isoscalar $S$-wave pion-nucleon interaction
in nuclear medium, though very weak, is more repulsive
than in vacuum \cite{FG,B0exp}.
Noting that the  leading
Weinberg-Tomozawa term vanishes and the pion-nucleon sigma term is
partly canceled by the effective range term,
higher order corrections to the one-body contribution is
substantial \cite{Meissner},
and also the two-body contribution is expected to play a significant
role;
this latter aspect has
been considered in explaining the threshold $S$-wave pion-nucleus
interaction \cite{FG,EW,MN}.

There are approaches for the effective pion masses
based on chiral symmetry \cite{WBW,TW}.
In this letter
we reexamine the problem in symmetric nuclear matter
in the framework of heavy baryon chiral perturbation theory.
All the contributions up to ${\cal O}(Q^7)$
and the one-body ${\cal O}(Q^8)$ contribution
in the pion self-energy will be calculated.
In addition, we will discuss some subtleties
of the problem
related to the definition of the pion fields.

The relevant quantity for the
effective pion mass $m_{\pi}^*$ in nuclear matter
we are interested in is the
in-medium self-energy of pion,
$\Pi(\omega^2)$,
which is defined with the pion momentum
$q^\mu=(\omega,\,{\vec 0})$.
To evaluate the self-energy systematically,
we employ Weinberg's formalism \cite{Weinberg} of heavy baryon
chiral perturbation theory (HBChPT), which has been quite successful
in understanding threshold pion-nuclear
interactions \cite{Weinberg,Kolck,PMR,PJM}.
In Weinberg's scheme,
Feynman diagrams are characterized by the order $Q^\nu$,
where $Q$ is the typical size of the momenta involved and/or
pion mass, which is regarded as small compared to the chiral
scale $\Lambda\sim 1$ GeV.
In this scheme, the Fermi momentum $k_F$ is thus counted as
of order $Q$.
For diagrams for the pion self-energy,
the chiral index $\nu$ is given as
\be
\nu= 2 + 2 L + \sum_i \nu_i,
\ \ \
\nu_i\equiv d_i + \frac{n_i}{2} -2
\ee
where $L$ is the number of the loop,
$d_i$ is the
the number of derivatives and/or pion masses,
and $n_i$ is the number of nucleon fields
attached to the $i$-th vertex.
In HBChPT, the Lagrangian consists of pions and nucleons,
while all other degrees of freedom are integrated out,
\begin{equation}
{\cal L}_{eff} = {\cal L}_0 + {\cal L}_1 +{\cal L}_2 +\cdots,
\end{equation}
where the subscripts are the chiral index $\nu_i$.
Terms relevant to computing the $S$-wave pion self-energy are
\bea
\Lag_0 &=& {\bar B}\left[ i v\cdot D
+ 2 i g_A S\cdot \Delta \right] B
- \frac{1}{2} \sum_A C_A \left({\bar B} \Gamma_A B\right)^2
+\, f_\pi^2 {\rm Tr}\left(i \Delta^\mu i \Delta_\mu
+ \frac{\chi_+}{4} \right),
\label{L0}\\
\Lag_1 &=& {\bar B} \Big[
 \frac{v^\mu v^\nu - g^{\mu\nu}}{2 m_N} D_\mu D_\nu
 +\frac{g_A}{m_N}\left\{S\cdot D, v\cdot \Delta\right\}
 \nonumber \\
 &&+\ c_1{\rm Tr} \chi_+
  +4\left (c_2-\frac{g_A^2}{8m_N}\right )\left (v\cdot i\Delta\right )^2
    +4c_3 i\Delta\cdot i\Delta \Big]B
\label{L1} \eea
with
$D_\mu = \frac{1}{2} \left( \xi_R^\dagger \del_\mu \xi_R
 + \xi_L^\dagger \del_\mu \xi_L\right)$,
$\Delta_\mu = \frac{1}{2} \left( \xi_R^\dagger \del_\mu \xi_R
 - \xi_L^\dagger \del_\mu \xi_L\right)$,
$\chi_{+} = \xi_R^\dagger \chi \xi_L + \xi_L^\dagger \chi^\dagger \xi_R$,
$\chi=m_\pi^2$, and
\be
\xi_R= \xi_L^\dagger =
  {\rm exp}\left(i\frac{{\vec \tau}\cdot {\vec \pi}}{2 f_\pi}\right).
\label{canon}\ee
The low-energy constants $c_1$, $c_2$ and $c_3$ are to be fixed from
experimental data; at present there are
several sets of values for the parameters determined
by fitting
low-energy pion-nucleon data\cite{Meissner}:
\bea
(c_1,c_2,c_3) &=& (-1.27\pm 0.12, 3.23\pm 0.19, -5.93\pm 0.08)\ \mbox{GeV}^{-1}
\;\;\mbox{(Fit1)},\nonumber\\
(c_1,c_2,c_3) &=& (-1.47\pm 0.09,3.21\pm 0.11,-6.00\pm 0.03)\ \mbox{GeV}^{-1}
\;\;\mbox{(Fit2)},\nonumber\\
(c_1,c_2,c_3) &=& (-1.53\pm 0.18,3.22\pm 0.25, -6.19\pm 0.09)\ \mbox{GeV}^{-1}
\;\;\mbox{(Fit3)}.
\eea

Figure 1 shows the nuclear diagrams we compute for the pion self-energy.
\begin{figure}[ht] %
\begin{center}
\epsfig{file=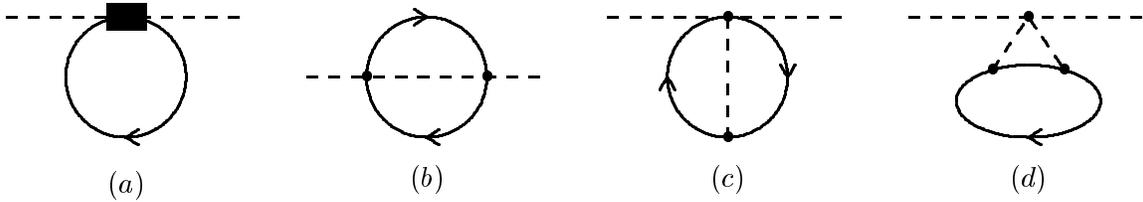}
\caption{\protect \small
Feynman diagrams that contribute to the pion self-energy.
Solid lines are nucleons, and dashed lines are pions.
The rectangle in $(a)$ denotes the whole set of diagrams
(including loop corrections) that contribute to the
$\pi N$ scattering amplitude.}
\label{something}
\end{center}
\end{figure}
Note that nucleon loops in {\em free} space are forbidden
in HBChPT, and all the nucleon lines (in solid line) drawn in Fig.~1
run only up to $k_F$, the Fermi momentum.
Throughout this work,
we employ the Fermi gas approximation for the wavefunction
of nuclear matter.
We acknowledge that
the diagrams in Fig.~1 have been studied in chiral models since long time
ago, 
see, for example, Ref. \cite{liter}.

The leading order (LO) contribution to the $S$-wave pion self-energy is
from the one-loop graph with the
Weinberg-Tomozawa term coming from ${\cal L}_0$,
which vanishes for symmetric nuclear matter.
If such a term survives, it would be \order4.
The next-to-leading order (NLO) contribution comes from
the one-loop graph with the
$\pi\pi NN$ vertices from ${\cal L}_1$,
which is nothing but the pion-nucleon sigma term.
Before going further, we note that all the diagrams
with a one-nucleon-loop only, called one-body contributions, are related to the pion-nucleon
scattering amplitude.
Particularly, all the contributions at $\omega=m_\pi$
can be replaced by the isoscalar $\pi N$ scattering length, $a_{\pi N}$:
\be
\Pi_{\rm 1-body}(\omega^2=m_\pi^2)=
 - 4\pi (1+\mu) a_{\pi N}
 \frac {2k_F^3}{3\pi^2}
\label{Pi1B}\ee
where $\mu\equiv \frac{m_\pi}{m_N}$.
Note that the above equation is correct to all order,
as far as the one-body contributions are concerned.
The scattering length has been calculated within HBChPT
\cite{beane},
\be
4\pi (1+\mu) a_{\pi N} = - \frac{m_\pi^2}{f_\pi^2} (4 c_1 - 4 c_{23})
 + \frac{3 g_A^2 m_\pi^3}{64 \pi f_\pi^4}
 + \cdots
\label{apiN}\ee
with $c_{23} \equiv \frac12 \left(c_2+c_3-\frac{g_A^2}{8m_N}\right)$.
In the {\it r.h.s.} of the above equation,
the first term is from NLO,
the second from the next-to-next-to-leading order (NNLO),
and the ellipsis denotes higher-order contributions.
At NNLO,
we also have genuine two-body contributions
which are drawn in Fig. 1($b-d$).
Including all of them, the pion self-energy
up to NNLO or up to \order6
reads
\bea
\Pi(\omega^2)&=&
 - \left[4\pi (1+\mu) a_{\pi N}
- \frac{\omega^2}{8\pi f_\pi^4} \sqrt{m_\pi^2-\omega^2}
 + (w^2-m_\pi^2) \left( \frac{4 c_{23}}{f_\pi^2} +
 \zeta \frac{3 g_A^2 m_\pi }{64 \pi f_\pi^4} \right)
  \right] \frac {2k_F^3}{3\pi^2}
\nonumber \\
&+&
 \frac{1}{16\pi^4 f_{\pi}^4}
 \left [2\omega^2 I_F(\sqrt{m_{\pi}^2-\omega^2})
  -\frac{g_A^2}{2} \left(m_{\pi}^2+ \zeta (w^2-m_\pi^2)\right)
  J_F(m_\pi) \right],
\label{pionself2}
\eea
where
\bea
I_F(m) &\equiv&
k_F^4-\frac{k_F^2m^2}{6}+\left (\frac{k_F^2m^2}{2}
 +\frac{m^4}{24}\right)\log\frac{4k_F^2+m^2}{m^2}
  -\frac{4}{3}mk_F^3\tan^{-1}\frac{2k_F}{m},
\nonumber \\
J_F(m) &\equiv&
k_F^4-\frac{k_F^2 m^2}{2}+\left (k_F^2 m^2
 +\frac{m^4}{8}\right)\log\frac{4k_F^2+m^2}{m^2}
  -2 m k_F^3\tan^{-1}\frac{2k_F}{m},
\label{IJ}\eea
and $\zeta=\frac{4}{3}$ for the pion field defined by eq.(\ref{canon}).
Here, $\zeta$ is an off-shell parameter
(that is effective only when the pion is off-shell),
which depends on the definition of the pion field.
For example, the canonical pion field introduced by
Weinberg \cite{Weinberg,wein68}\footnote{\protect
The covariant derivative of the pion field we are referring to
has the form
$\del_\mu {\vec \pi}/(1+{\vec \pi}^2/(4 f_\pi^2))$.}
corresponds to $\zeta= 3$.
In fact,
the value of $\zeta$ is not fixed
by chiral symmetry
and can be any arbitrary number \cite{wein68}.
As another interesting definition of the pion field,
we discuss here the background field method (BFM) \cite{ecker}.
In BFM, the covariant quantities are defined as
\be
\xi_R= \exp\left(i\frac{{\vec \tau}\cdot {\vec \pi}_{\rm cl}}{2 f_\pi}\right)
 \exp\left(i\frac{{\vec \tau}\cdot {\vec \pi}_{\rm fl}}{2 f_\pi}\right),
\ \ \
\xi_L= \exp\left(-i\frac{{\vec \tau}\cdot {\vec \pi}_{\rm cl}}{2 f_\pi}\right)
 \exp\left(-i\frac{{\vec \tau}\cdot {\vec \pi}_{\rm fl}}{2 f_\pi}\right),
\label{bfm}\ee
where ${\vec \pi}_{\rm cl}$ is the pion field that appears as external legs,
while the ${\vec \pi}_{\rm fl}$ is for the internal pion lines.
Often the
BFM removes some of artificial contributions
and simplifies the calculation.
In our case, the contribution of Fig. $(1c)$ vanishes with BFM;
with the canonical pion field of eq.(\ref{canon}), Fig. $(1c)$ is not zero
but canceled out by a part of the contribution of Fig. $(1d)$.
For physical observables where all the external particles
are on mass shell, however,
BFM gives exactly the same results
as any other definition of pion field does.
The value of $\zeta$ obtained  by using BFM is $\zeta=4$.
\label{pion self-energy}

\begin{figure}[htbp]
\epsfig{file=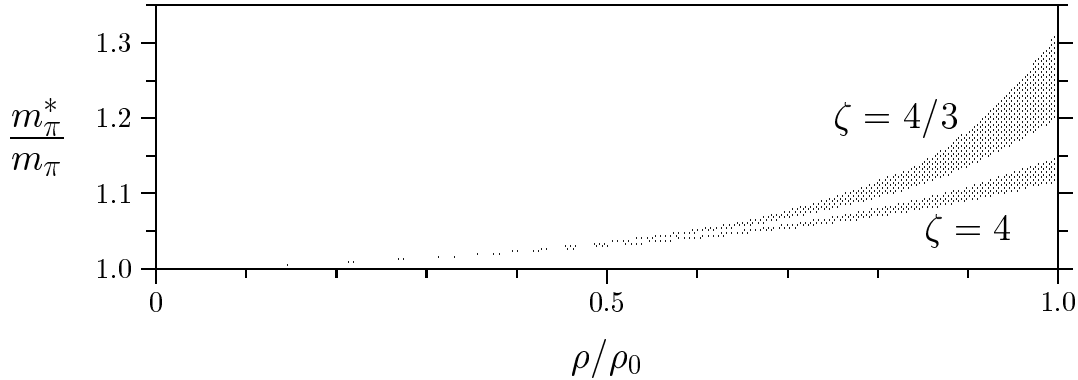}
\caption[U]{\protect\small
The ratio $m_\pi^*/m_\pi$ as a function of $\rho/\rho_0$.
The upper shade is
obtained with the canonical $U$-field ($\zeta=\frac43$),
while the lower shade with the BFM ($\zeta=4$).
In each shade, 
the upper (lower) bound corresponds to Fit 1 (Fit 3).
Fit 2 gives almost the same results as Fit 3.}
\end{figure}

Now we are in a position to discuss the in-medium pion mass $m_\pi^*$.
The conventional way of determining the in-medium mass is
to find the zero of the
inverse of the pion propagator,
$w^2-m_\pi^2-\Pi(w^2)$,
or
\be
\mpis^2= m_\pi^2 + \Pi(\mpis^2).
\label{mpistar}\ee
But since $\Pi(\mpis^2)$ with $\mpis\neq m_\pi$ depends on $\zeta$,
the resulting $\mpis$ also depends on the off-shell parameter $\zeta$.
The results are depicted in Fig. 2
with two popular choices of $\zeta$;
the upper shade is for the case with
$\zeta= 4/3$ (the canonical field of eq.(\ref{canon})),
and the lower shade is for
$\zeta=4$ (BFM).
The results with $\zeta=3$ lie in between the two.
In each case, the parameter set Fit 1 gives the biggest $m_\pi^*$,
and Fit 3 results in the lowest value.
At normal nuclear matter density,
the ratio $m_\pi^*/m_\pi$ is
$1.20 -1.32$ with $\zeta=4/3$,
and $1.12 - 1.15$ with $\zeta=4$.
Thus, apart from the non-trivial dependence
on the parameter set of low-energy constants,
there is quite substantial $\zeta$-dependence.
We note that in the chiral limit $m_\pi=0$, however,
$\mpis$ remains zero in
all the cases.

An interesting observation is that, when the nuclear
density reaches around $\rho= (1.3 \sim 1.7)\ \rho_0$ with
$\zeta=(4/3\sim 4)$, where $\rho_0$ is the normal nuclear matter density,
the in-medium pion mass decreases abruptly below the
pion mass in the vacuum.
This is caused by the fact that when the density
increases, the value of the $\Pi'(\omega^2)\equiv \frac{\del}{\del
\omega^2} \Pi(\omega^2)$ is close to one 
and makes the solution of eq.(\ref{mpistar}) unstable
at the density mentioned above.

The off-shell ambiguity may indicate that
the effective pion mass defined by eq.(\ref{mpistar})
is not a physical observable,
which should be independent of the definition of the pion field.
Indeed the effective pion mass is measured only indirectly in experiments
with the help of optical potentials.
A thorough study on this subject, which should be quite important,
has not been done in this work
and will appear in a separate work.

In this letter, 
however, we introduce a new way 
to extract the effective pion mass unambiguously.
So far the
pion self energy has been obtained to a certain order $\nu$, and
then treated non-perturbatively in obtaining the effective mass
through eq.(\ref{mpistar}). 
To avoid the above mentioned ambiguities, we propose to get
the effective mass itself up to order $\nu$.
The new procedure gives us exactly the same results
as the previous method up to $\nu$, all the differences residing
in higher orders. 
Particularly if the gap $(\mpis^2-m_\pi^2)$ is
of order of $Q^2$ as the naive counting indicates, 
there is absolutely no difference. 
In actual cases, the gap is much smaller
than in the naive counting,
\be
\mpis^2 - m_\pi^2 = \order5,
\label{mpiorder}\ee
since the non-vanishing-leading-order (or NLO)
of the pion self-energy is \order5{}.
Inserting this into eq.(\ref{pionself2}) with $\omega^2=\mpis^2$,
we see that
{\em all the $\zeta$-dependent terms
in $\Pi(\mpis^2)$ are of ${\cal O}(Q^9)$},
which is beyond the scope of this work and will be neglected.
In other words,
there is no off-shell ambiguity in
$\Pi(\mpis^2)$ up to ${\cal O}(Q^8)$.
Furthermore, $\Pi(\mpis^2)$ is the same as $\Pi(m_\pi^2)$
up to \nlo7, that is,
$\Pi(\mpis^2)=\Pi(m_\pi^2) + \nlo8$.
Expanding $\Pi(\mpis^2)$ based on our
``detailed'' power counting rule,
\be
\Pi(\mpis^2) = \sum_{\nu=5}^\infty \Pi^{(\nu)}(\mpis^2),
\ee
our two-loop order calculation determines
the first two leading order contributions unambiguously:
\bea
\Pi^{(5)}(\mpis^2)&=&
 - 4\pi (1+\mu) a_{\pi N} \frac{2k_F^3}{3\pi^2},
\label{Pi5}\\
\Pi^{(6)}(\mpis^2)&=&
 \frac{\mpi^2}{16\pi^4 f_{\pi}^4}
 \left [2 k_F^4 -\frac{g_A^2}{2} J_F(m_\pi) \right].
\label{Pi6}
\eea
We remark that
$\Pi^{(5)}(\mpis^2)$ in eq.(\ref{Pi5}) includes
all the one-body contributions
at $\omega^2=m_\pi^2$ up to infinite order,
since the experimental scattering length $a_{\pi N}$ contains
all the higher order contributions
as well as the leading \order2 contribution.
We can also show that there is no ${\cal O}(Q^7)$
contribution,
\bea
\Pi^{(7)}(\mpis^2)&=& 0,
\label{Pi7}
\eea
by the following arguments.
All the one-body contributions are already absorbed
into $\Pi^{(5)}$ at $\omega^2=m_\pi^2$,
and the difference
$\Pi(\mpis^2)-\Pi(m_\pi^2)$ is \nlo8,
which can be read from eq.(\ref{pionself2}).
The two-body contributions at \nlo7 are
two-loop diagrams which contain one vertex from ${\cal L}_1$;
however, they
are all isovector and therefore cannot contribute to the
pion self-energy in symmetric nuclear matter.
Contributions involving three or more nucleons
begin to appear only at \nlo8.

At \nlo8
there are one-, two- and three-body contributions,
\be
\Pi^{(8)}(\mpis^2) =
\Pi^{(8)}_{\rm 1-body}(\mpis^2)
+\Pi^{(8)}_{\rm 2-body}(\mpis^2)
+\Pi^{(8)}_{\rm 3-body}(\mpis^2).
\ee
The one-body contribution at this order
comes from
the $\omega$-dependence in $\Pi(\omega^2)$,
\bea
\Pi^{(8)}_{\rm 1-body}(\mpis^2)&=&
  4\pi (1+\mu) a_{\pi N} \,
  \frac{4 c_{23}}{f_\pi^2} \,
   \left(\frac {2k_F^3}{3\pi^2}\right)^2.
\label{Pi8}\eea
The two- and three-body contributions at this order
are not discussed in this work.
\begin{figure}[htbp]
\epsfig{file=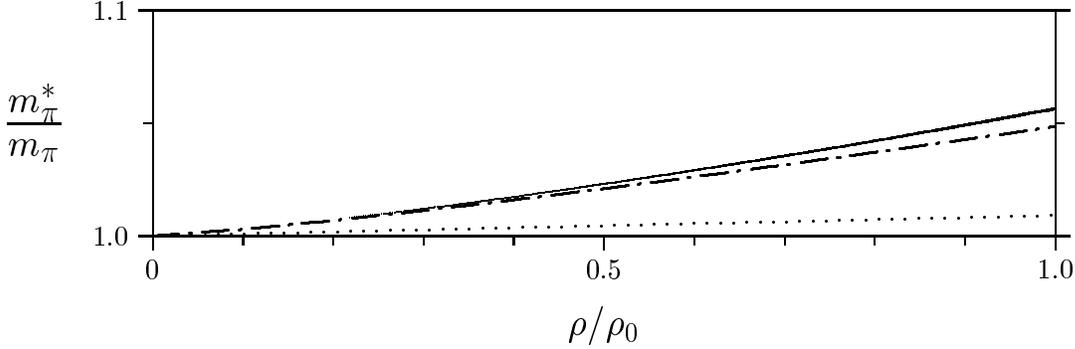}
\caption[U]{\protect\small
The cumulative graphs for
$m_\pi^*/m_\pi$ vs. $\rho/\rho_0$.
The dotted, dotted-dashed and solid lines stand for the
results up to \nlo5, \nlo6 and \nlo8.
The result up to \nlo8 does not contain
the two- and three-body contributions at this order,
see the explanation in the text.
}
\end{figure}

The numerical results are given in Fig. 3.
The \nlo5 (dotted line) contribution is tiny because of the smallness of the
$\pi-N$ scattering length.
The two-body contribution embodied in \nlo6 (dotted-dashed line)
gives the main contribution,
about 6\ \% increase
of the in-medium pion mass
at normal nuclear matter density.
As mentioned above, there is no \nlo7 contribution.
Just for discussion,
we have drawn \nlo8 results (solid line)
without the two- and three-body contributions.
This (partial) \nlo8 contribution at this order is not significant, which is
again due to the smallness of the $\pi-N$ scattering length.
For the same reason, the dependence on the parameter set up to this order is tiny and
invisible in the figure.

By expanding $\Pi(\mpis^2)$ with the {\em detailed} power
counting rule which considers the gap $(\mpis^2-m_\pi^2)$ to be of order $Q^5$,
we find that
the in-medium pion mass
increases only very mildly as the nuclear matter density increases.
At normal nuclear density, $\mpis$ is larger than $\mpi$
only by $(5\sim 6)\ \%$.
This result has little dependence on the parameter set used,
and is independent of the off-shell parameter $\zeta$.
Our results support the phenomenological study given in
Ref.\cite{WBW}.
The complete calculation up to \nlo8 is feasible and involves
computing three-body contributions as well as one- and two-body
contributions, and it would play more important roles
in higher density regions.

Let us make concluding remarks. 
We have considered two schemes for computing the effective pion mass 
in symmetric nuclear matter based on chiral perturbation theory:
a navie power counting scheme (eqs.(\ref{Pi1B}-\ref{IJ}, \ref{mpistar}), Fig. 2)
and a more detailed power counting scheme (eqs.(15-19), Fig. 3).
The discrepancies in $\mpis^2$ are of order of ${\cal O}(Q^9)$ 
or higher order. 
We believe the latter is more meaningful than the former,
because the latter is independent of the offshell parameter $\zeta$.
While  of higher order than considered,
the possible dependence on the
off-shell parameter at ${\cal O}(Q^9)$ remains to be addressed.
Especially the radical differences between the two schemes
(Fig. 2 and Fig. 3)
are quite alarming. Thus our results up to NNLO should
be taken with caution. To be more definite, 
a calculation up-to-${\cal O}(Q^9)$
would be required.

In overcoming the problem caused by the off-shell ambiguity, 
the idea of Brown-Rho (BR) scaling \cite{BR}
may shed some light on this problem.
If we follow the BR scaling, the the parameters of
the pion self-energy should be replaced by the in-medium effective values.
Let $\Pi^*(w^2)$ be the BR scaled pion self-energy.
Then the equation for the $m_\pi^*$ reads
\be
\mpis^2= m_\pi^2 + \Pi^*(\mpis^2).
\ee
with
\bea
\Pi^*(\mpis^2)&=&
 - 4\pi (1+\mu) a_{\pi N}^* \frac {2k_F^3}{3\pi^2}
+
 \frac{\mpis^2}{16\pi^4 {f_\pi^*}^4}
 \left [2 k_F^4
  -\frac{{g_A^*}^2}{2}
  J_F(\mpis) \right].
\label{pionselfBR}
\eea
We note that there is no $\zeta$-dependence,
and these equations are essentially the same as eqs.(\ref{Pi5}-\ref{Pi8}) 
except that all the parameters
are replaced by their in-medium counterparts.
This is because in nuclear medium
the effective on-shell condition is $\omega=\mpis$
and  $\Pi^*(\mpis^2)$ is effectively an on-shell quantity.
However, 
in order to perform a numerical analysis,
we need the value of $a_{\pi N}^*$ 
as well as other effective
parameters,
which requires more studies.

On finalizing this work, 
we have learned of the existence of a similar work
by Kaiser and Weise\cite{weise}.
Their work is more extensive than ours in that
it includes
pions and kaons in isospin asymmetric 
nuclear matter 
but
the effective pion mass is simply assumed to be
$\mpis^2=m_\pi^2 + \Pi(m_\pi^2)$.

\vspace{1 in} \centerline{\bf ACKNOWLEDGMENTS} \vspace{0.3 in} We
are grateful to Jerry Miller and Mannque Rho for valuable
comments. 
We thank E. Oset and N. Kaiser for comments.
TSP is indebted to S.-i. Ando, F. Myhrer and K. Kubodera
for discussions.
H. J. acknowledges the hospitality
of the nuclear theory group of University of Washington.
This work was supported in part by Korea Research
Foundation (KRF-grant-1999-015-DI0023) and in part by Korea
Science and Engineering Foundation (1999-2-111-005-5). The work of
H. J. was also supported in part by Korea Ministry of Science and
Technology (99-B-WB-07-A-03).

\end{document}